\documentclass[twocolumn]{aastex63}

\newcommand{\ktwo}{\textit{K2}\,}

\usepackage{lipsum}
\usepackage{amsmath} 
\usepackage{rotating}
\shorttitle{Short-Timescale Hot Spot Variability}
\shortauthors{Biddle et al.}

\begin{document}

\title{Amplitude Modulation of Short-Timescale Hot Spot Variability}

\correspondingauthor{Lauren I. Biddle}
\email{lbiddle@lowell.edu}

\author[0000-0003-2646-3727]{Lauren I. Biddle}
\affiliation{Lowell Observatory, 1400 W. Mars Hill Rd. Flagstaff. Arizona. 86001. USA.}
\affiliation{Department of Physics \& Astronomy, Northern Arizona University, 527 S. Beaver St. Flagstaff. Arizona. 86011. USA.}

\author[0000-0003-4450-0368]{Joe Llama}
\affiliation{Lowell Observatory, 1400 W. Mars Hill Rd. Flagstaff. Arizona. 86001. USA.}

\author[0000-0002-8863-7828]{Andrew Cameron}
\affiliation{SUPA, School of Physics and Astronomy, North Haugh, St Andrews, Fife, KY16 9SS, UK}

\author[0000-0001-7998-226X]{L. Prato}
\affiliation{Lowell Observatory, 1400 W. Mars Hill Rd. Flagstaff. Arizona. 86001. USA.}

\author[0000-0002-1466-5236]{Moira Jardine}
\affiliation{SUPA, School of Physics and Astronomy, North Haugh, St Andrews, Fife, KY16 9SS, UK}

\author[0000-0002-8828-6386]{Christopher M. Johns-Krull}
\affiliation{Department of Physics \& Astronomy, Rice University, 6100 Main St. MS-108, Houston, TX 77005}



\begin{abstract}
Variability of Classical T Tauri Systems (CTTS) occurs over a vast range of timescales. CTTS in particular are subject to variability caused by accretion shocks which can occur stochastically, periodically, or quasi-periodically on timescales over a few days. The detectability of young planets within these systems is likely hampered by activity; therefore, it is essential that we understand the origin of young star variability over a range of timescales to help disentangle stellar activity from signatures of planetary origin. We present analysis of the stochastic small-amplitude photometric variability in the \ktwo lightcurve of CI Tau occurring on timescales of $\la$1 d. We find the amplitude of this variability exhibits the same periodic signatures as detected in the large-amplitude variability, indicating that the physical mechanism modulating these brightness features is the same. The periods detected are also in agreement with the rotation period of the star ($\sim$6.6 d), and orbital period of the planet ($\sim$9.0 d) known to drive pulsed accretion onto the star.

\end{abstract}

\section{Introduction} \label{sec:intro}

Young low mass stars with protoplanetary disks, otherwise referred to as Classical T Tauri stars (CTTS), are known to be highly variable over a range of timescales. CTTS have been observed to exhibit long-term, consistent rotation periods ranging from days to weeks, identified through flux modulation caused by cool spots with lifetimes lasting months to years \citep{Stelzer2003,Herbst2007,Carvalho2020}. Evidence for magnetic activity cycles similar to those in the Sun have also been observed on these young stars \citep{Cohen2004}. CTTS also undergo significant brightness changes on short timescales from minutes to hours to days, varying in both strength and temporal profile depending on the source. Flares account for some of this variability; young, low-mass stars are fully convective and hence manifest the most powerful flares \citep{Feinstein2020}. Higher flare rates are seen in cooler stars (T$_{\textrm{eff}}\leq$ 4000 K). Additionally, CTTS host actively accreting circumstellar disks, in which disk material travels along the magnetic field lines connecting the disk to the star. The accreting material reaches supersonic free-fall speeds, creating a shock on the star at the foot of the accretion stream \citep{Koenigl1991,Calvet1992,Shu1994}. The resulting flux increase seen in the photometric lightcurve points to a hot spot. The overall increase in brightness from a hot spot can last as long as there is continuously accreting material, but rapidly changing accretion rates can cause the shock to produce sporadic, short bursts which can vary in strength and duration from minutes to hours depending on the rate of change of accretion \citep{Robinson2017}.

A consequence of the substantial variability in young stars is that it greatly hinders the detectability of young planets around them. In terms of RV detection, cool spots on the rotating surface of an inclined star that would be visible at all times produce a variable signature that closely resembles a planet-induced RV signal and can lead to a false-positive detection \citep{Queloz2001,Huerta2008}. CTTS variability can also severely limit planet detectability in terms of planetary transits especially in stars actively undergoing accretion. Accretion hot spots can increase the system luminosity by several hundred percent \citep{Herbst1994}, making it more difficult to detect a transiting planet within the total observed system brightness; the additional luminosity from hot spots will dilute the transit, and the accretion shock variability can obscure it almost entirely. Directly imaging young planets also has its challenges. The nearest star-forming regions are relatively distant ($>$100 pc), limiting our ability to reliably image planetary mass companions. 

Because of these difficulties, only a handful of planets around young stars have been detected (including but not limited to 2M1207 b \citealt{Chauvin2004, Chauvin2005}; LkCa 15 b \citealt{kraus2012}; CI Tau b \citealt{Johns-Krull2016}; V830 Tau b \citealt{donati2017}; K2-33 b \citealt{mann2016,david2016}; Tap 26 b \citealt{yu2017}; and V1298 Tau b,c,d,e \citealt{David2019a,David2019b}), though a few have been called into question (e.g., LkCa 15 b, \citealt{Currie2019}; V830 Tau b, \citealt{Damasso2020}). To date, CI Tau b is the only hot Jupiter detected around a CTTS. Under the assumption that the disk is aligned with the equatorial plane of the star, the inclination of the CI Tau system acquired with ALMA is $\sim$49 degrees \citep{Clarke2018}, indicating that the planet is non-transiting. However, a period-search of CI Tau's \ktwo lightcurve identified a periodic signal at $\sim$9.0 d \citep{Biddle2018}, consistent with the orbital period of the $\sim$11.0 Jupiter-mass planet found using RV measurements \citep{Johns-Krull2016}, located just within the co-rotation radius at the inner edge of the disk ($\sim$0.1 au). The basis of such a detection relies on the interactions between the planet and accreting disk material, producing a periodic signal on the timescale of the planet's orbit. This is the first instance where an accretion signature itself has been demonstrated to trace the presence of a young planet. Understanding the characteristic signatures of accretion is therefore paramount to identifying additional young star systems whose photometric time series could suggest interactions of this nature. 

To that end, we build upon the analysis presented in \citet{Biddle2018}; here we examine the \ktwo light curve of CI Tau focusing on the extreme short-term photometric variability ($\lesssim$1 d) and its relation to accretion processes. The unprecedented sampling rate and precision of \ktwo offers the opportunity to characterize variability of young stars in great detail.

\section{Data and Analysis} \label{sec:2}

NASA's \textit{Kepler} Spacecraft acquired long-cadence time-series photometry of CI Tau (EPIC 247584113) during the \ktwo mission Campaign 13 between 8 March 2017 and 27 May 27 2017 UTC. We acquired the lightcurve output from the The Pre-search Data Conditioning Simple Aperture Photometry (PDCSAP) pipeline \citep{jenkins2010}. The PDCSAP pipeline produces lightcurves from Single Aperture Photometry (SAP) data products out of the Kepler data processing pipeline from which systematic trends are removed by the PDC (Pre-search Data Conditioning) method \citep{Twicken2010}. The PDC is known to remove trends with periodicities greater than $\sim$10 days, including those which could be astrophysical in nature. To verify that signals on this timescale were maintained, we cross-referenced periodic signatures of the PDCSAP output to those of the SAP output (which retains long-term trends) and found no significant differences between the two. We then applied the K2SC detrending algorithm \citep{aigrain2015,aigrain2016} to the PDCSAP lightcurve. The K2SC algorithm is designed to be able to preserve the astrophysical variability of the star while removing additional position-dependent systematics in the data. As was the case with the PDCSAP output, the resulting lightcurve shows no indication that periodicities on days-long timescales were detrended from the data.

\begin{figure*}[ht!]
\centering
    \includegraphics[width=1\textwidth]{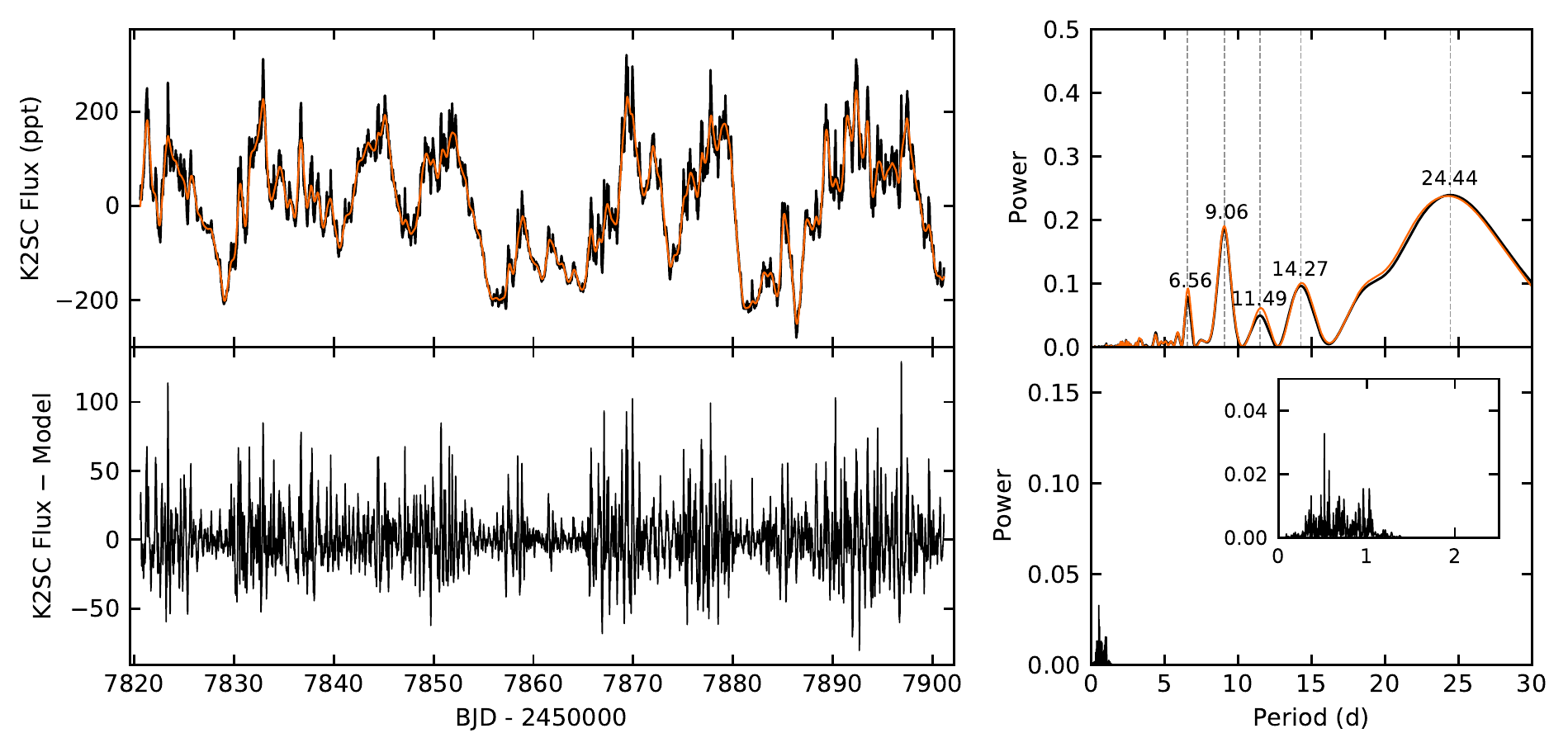}
    \caption{\textit{Upper Left:} A comparison of the CI~Tau lightcurve detrended with K2SC (black) and the Gaussian Process model (orange). \textit{Upper Right:} The Lomb-Scargle periodograms of the K2SC-detrended lightcurve and the Gaussian Process model in black and orange, respectively. \textit{Lower Left:} The residual signal from the subtraction of the Gaussian process model from the K2SC lightcurve. \textit{Lower Right:} The Lomb-Scargle periodogram analysis of the residual signal. The inset panel provides a zoomed-in view of the periodogram results.}
\label{fig:lc}
\end{figure*}

\begin{figure*}[ht!]
\centering
    \includegraphics[width=1\textwidth]{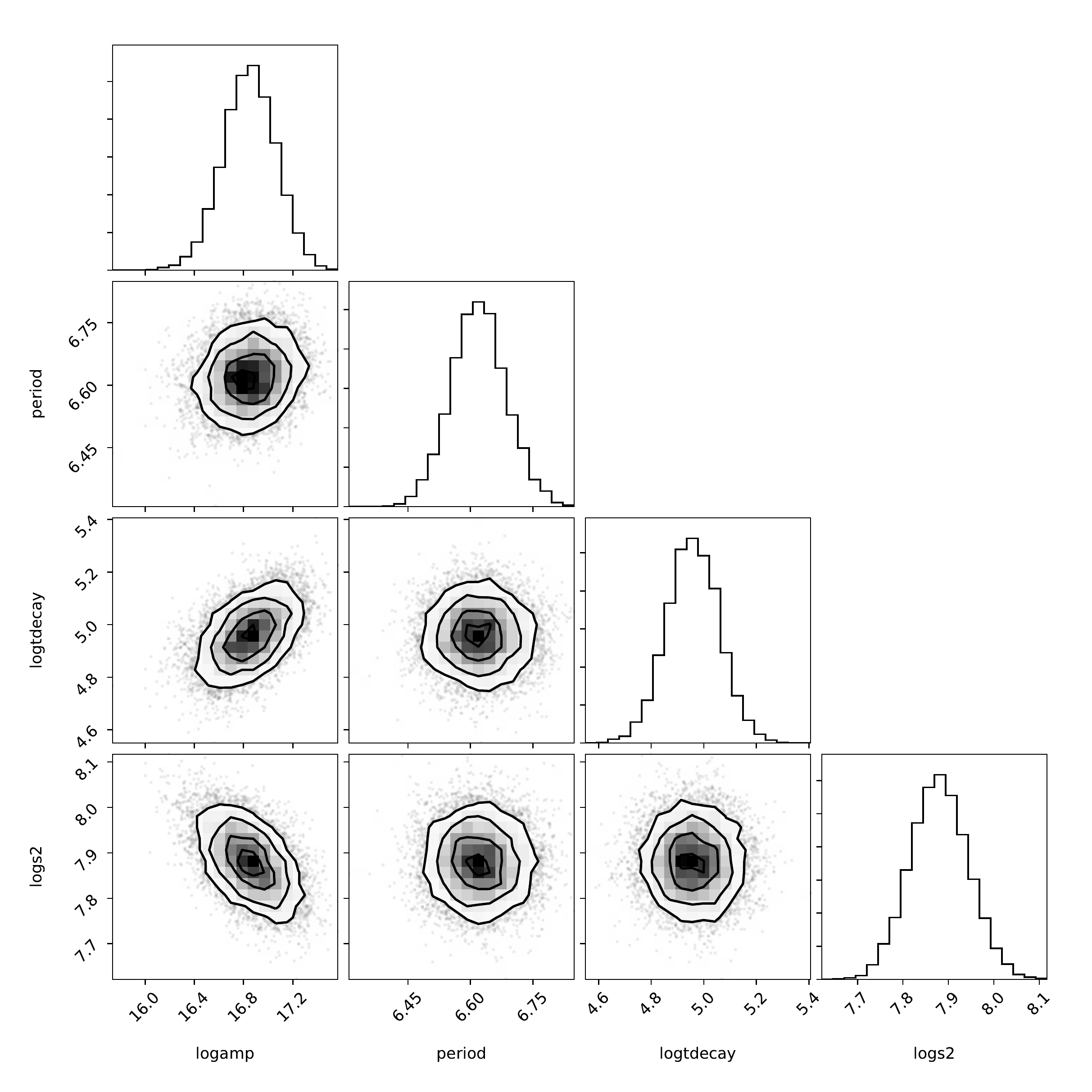}
    \caption{The posterior constraints on the hyperparameters of the GP model (see Appendix \ref{sec:appendix}). The black contours contain the 0.5,  1,  1.5,  and 2-sigma credible regions in the marginalized planes. The histograms along the diagonal show the posterior for each parameter.}
\label{fig:corner}
\end{figure*}

We isolated the small-scale variability by modeling and subtracting out the larger periodic and quasi-periodic variability that occurs over longer timescales greater than a few days. The model is computed at every timestamp in the K2 lightcurve using the Gaussian Process (GP) framework known as \textsc{celerite} developed by \citet{Foreman-Mackey2017}. The framework requires the covariance function to be represented by a mixture of exponentials but does not require the data to be evenly spaced or to be associated with uniform uncertainties. We constructed a covariance function (Appendix \ref{sec:appendix}) to model the stellar rotation in addition to the pseudo-periodic variability caused by magnetic features such as faculae and plage on the stellar surface. We note that the covariance function produces results consistent with that of \citet{Foreman-Mackey2017}. The full set of parameters and their priors are provided in Table \ref{table:priors}. Following \citet{Foreman-Mackey2019}, we applied a Savitzky-Golay filter to the data and then sigma clipped the results. Figure \ref{fig:lc} shows our stellar rotation model applied to CI Tau's \ktwo lightcurve and Figure \ref{fig:corner} summarizes the posterior constraints of the model parameters. 

\begin{table}[]
\centering
\caption{Parameters and Priors for the Gaussian Process Model \label{table:priors}. $^\textrm{a}$ Rotation term informed by \citet{Biddle2018}. }
\begin{tabular}{cc}
\hline
Parameter                                      & Prior                                          \\ \hline
$\ln(\textrm{A}/\textrm{ppt})$                 & $\mathcal{N}$(9.48, 5)                         \\
$\ln(\textrm{P}/\textrm{days})$                & $\mathcal{N}(\ln(6.6^\textrm{a}),~0.01)$ \\
$\ln(\textrm{t}_\textrm{decay}/\textrm{days})$ & $\mathcal{N}$(5.0, 0.1)                        \\
$\ln(\textrm{s2})$                             & $\mathcal{N}$(22.0, 0.1 )                      \\ \hline
\end{tabular}
\label{table:priors}
\end{table}

For both the GP model and \ktwo lightcurve, we computed a generalized Lomb-Scargle periodogram \citep{Zechmeister2009} using the \textsc{python} AstroML\footnote{\url{http://www.astroml.org/index.html}} package. We compute the Lomb-Scargle periodogram for both the \ktwo lightcurve and GP model. Each periodogram searched 10,000 frequencies within a window consistent with Nyquist sampling \citep{Press1992}. The GP was able to capture the large-scale variability responsible for producing the periodic signals identified in the \ktwo lightcurve (Figure \ref{fig:lc}). We then subtracted the GP model from the \ktwo lightcurve leaving only the variability with amplitudes much smaller than that which produced the strongest signals in the Lomb-Scargle analyses. For consistency, we also computed the periodogram of the residuals. As expected, the results show no indication of periods identified from the large-scale variability prior to subtraction (Figure \ref{fig:lc}) or periodic artifacts that could arise from underfitting the large-scale light curve variability. 

\begin{figure*}[ht!]
    \centering
    \includegraphics[width=\textwidth]{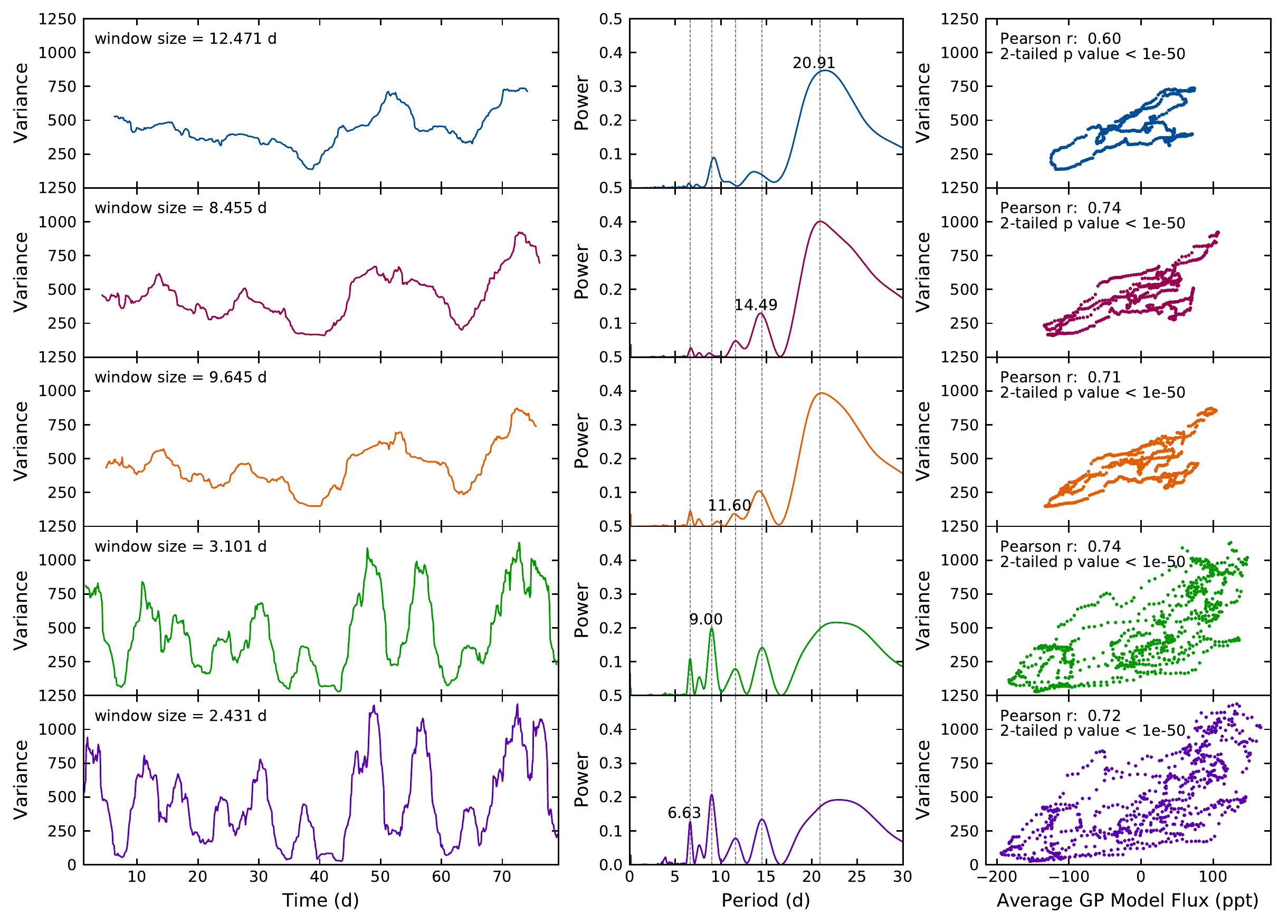}
    \caption{\textit{Left Column:} The variance curves calculated using the sliding window size associated with the respective peak periods identified in the two-dimensional Lomb-Scargle periodogram. \textit{Center Column:} Cross-cuts of the two-dimensional Lomb-Scargle periodogram corresponding to the window size used to calculate its respective variance curve. \textit{Right Column}: Correlation between the variance curve and the average flux of the GP model within the sliding window.}
    \label{fig:crosscuts}
\end{figure*}

The amplitude of the scatter exhibits time-dependent variability. We quantify the changing amplitude by computing the variance of the scatter over the duration of the observations. Each measure of the variance is calculated from the points that fall within the bounds of a sliding window. The window advances 0.1 days in time along the lightcurve, producing a ``variance curve'' (Figure \ref{fig:crosscuts}) with which we can perform a period search. We repeat this process for a range of window sizes to ensure that any periodicity that may be present is not suppressed by the size of the window itself. In total, we calculate 200 variance curves, each calculated using a window size between 0.20 days and 15 days, all of which were equally spaced within this range.

\begin{figure*}[ht!]
\centering
    \includegraphics[width=1\textwidth]{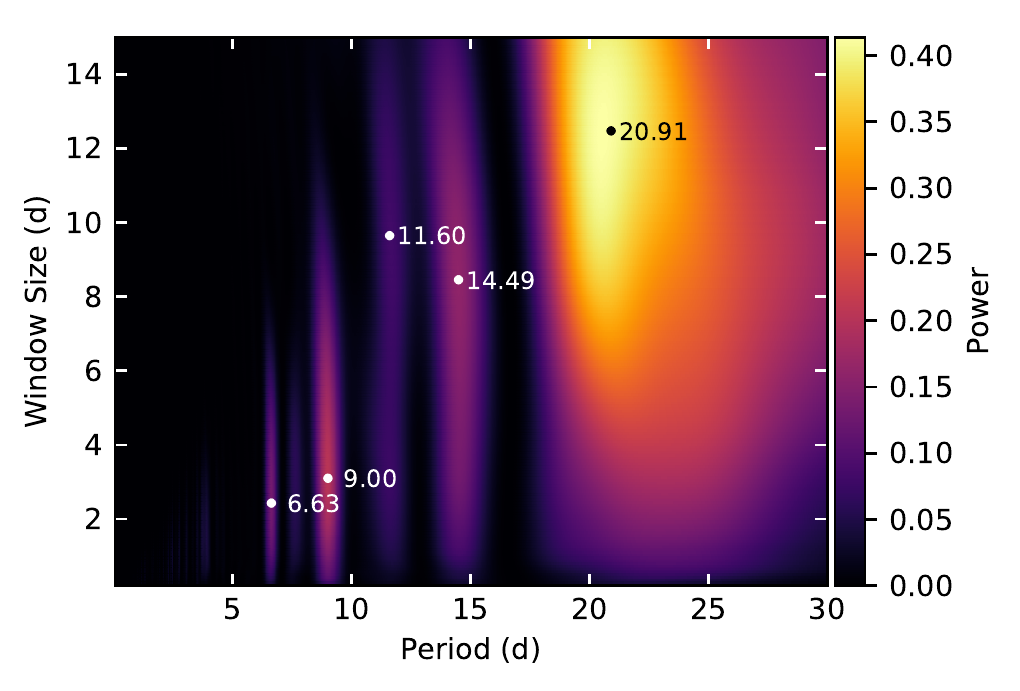}
    \caption{Two-dimensional Lomb-Scargle periodogram of the variance curve. The resulting periods are displayed adjacent to the point indicating the location of each respective peak.}
\label{fig:2Dperiodogram}
\end{figure*}

Figure \ref{fig:2Dperiodogram} maps the results of the period search performed on all variance curves. The plot reveals several peaks that closely resemble the primary peaks identified in the \ktwo lightcurve (Figure \ref{fig:lc}, \citealt{Biddle2018}), despite the clear removal of the original periodic components of that analysis. Peaks occur at 6.62$\pm$0.2 d, 9.00$\pm$0.5 d, 11.60$\pm$1.5 d, 14.49$\sim$1.7 d, and 20.91$\pm$4.4 d. We determine the uncertainty on the periods from the full width at half of the maximum of the power distribution surrounding each period \citep{Ivezic2014}. The contribution of the $\sim$6.6 d period peaked in the variance curve calculation using a window size of 1.911 d, the $\sim$9.0 d period contribution peaked with a window size of 2.506 d, and the $\sim$20.0 d period contribution peaked with a window size of 8.158 d. Using both the analytic solution \citep{Zechmeister2009} and a Monte Carlo bootstrap algorithm, we calculated the false alarm probability (FAP) of the periods using the variance curves that produce the respective peak signals. For all periods, both methods yield a FAP of $<$10$^{-6}$. Figure \ref{fig:crosscuts} shows a 1D representation of the variance curves corresponding to each peak's maximum contribution to the time-dependent variability of the amplitude of the scatter of the residuals.

We find that the amplitude of the small-scale variability shows a positive correlation with the brightness of the system flux (Figure \ref{fig:crosscuts}). We quantify this trend for each cut of the 2D periodogram by calculating the Pearson’s correlation coefficient, $r$, between the variance curve and the average flux of the GP model lightcurve within the sliding window used to calculate the variance curve. The resulting values of r range between 0.71 - 0.74 with the exception of the variance curve corresponding to the peak $\sim$20 d period with the lowest correlation coefficient, $r = 0.60$. All resulting two-tailed $p$ values are less then 10$^{-50}$.

\section{Discussion} \label{sec:3}

The similar occurrence in periodicity within both the large- and small-scale variability provides evidence that these signals originate from the same physical process, possibly related to accretion. The large-scale variability occurs as a result of the varying luminosity generated by accretion hotspots as they come into and out of view \citep{Koenigl1991}. In terms of small-amplitude variability, stochastic brightness fluctuations changing on a $\la$1 d timescale could be a characteristic observational signature associated with accretion shocks. Numerical simulations of accretion shocks predict rapid variability in the shock (e.g., \citealt{Orlando2010}). The small-scale variability occurs when shocks form at the base of the accretion column where hotspots are located. We therefore expect to see an increased occurrence of small-scale variability when the hotspots are in view. The observations are in agreement with this model. 

Additionally, the timescales by which the flux variations occur agree with what we can expect for the hot spot response time, which in this case is approximately the sound-crossing time, $t_{\textrm{cross}}$. If the spot is of linear scale, then for a hot spot that is $\sim$10 degrees in width, we can estimate $t_{\textrm{cross}}$ by
\begin{equation}
    t_{\textrm{cross}} \sim \frac{\pi R_{\star}}{18} \sqrt{\frac{m_{\textrm{p}}}{k_{\textrm{B}} T_\textrm{s}}},
\end{equation}
where $R_{\star}$ is the stellar radius, $k_{\textrm{B}}$ is the Boltzmann constant, $m_{\textrm{p}}$ is the mass of a proton, and $T_{\rm{s}}$ is the post-shock temperature, defined as
\begin{equation}
    T_\textrm{s} = \frac{2m_{\textrm{p}}}{k_{\textrm{B}}}\frac{\gamma - 1}{(\gamma + 1)^{2}}v_{\textrm{acc}}^{2}.
\end{equation}
Here, $v_{\textrm{acc}}$ is the velocity of the accreting material, which moves at approximately the free-fall speed ($\sim$430 km/s), and $\gamma$ is the polytropic index ($\gamma = \frac{5}{3}$). We calculate $T_{\rm{s}}\sim10^{6}$ K, resulting in $t_{\textrm{cross}} \sim$ 30 min, consistent with our observations. The resulting temperature also predicts time-correlated variations in X-ray emission, a possible motivator for high energy small satellite missions.

Evidence of rapid, small-amplitude changes in rates of accretion on a timescale of minutes to hours have been observed in spectra of CTTS \citep{Costigan2014}. Short-period, rapid changes in the accretion rate onto CTTS form as a result of variable density in the inner disk \citep{Robinson2017}. One source of density variations at the inner edge of the disk is turbulence (e.g., \citealt{Fromang2006}), which is likely to occur stochastically. Magnetorotational instabilities can cause turbulence, which can affect the amount of material available for accretion \citep{Romanova2012}. Another source of density changes includes massive collections of matter within the inner edge of the disk, which can perturb the density and stimulate turbulence. For example, a massive planet can perturb the density via planet-disk interactions \citep{Biddle2018}. A hot Jupiter like the one detected around CI Tau \citep{Johns-Krull2016} could trigger pulsed accretion at the inner edge of the disk \citep{Teyssandier2019}. The existence of a hot Jupiter in this system is supported by independent RV measurements producing an orbital period of $\sim$9.0 d \citep{Johns-Krull2016} as well as a recent detection of CO at the velocity corresponding to the RV period \citep{Flagg2019}, indicating the presence of a structure orbiting near the star's co-rotation radius.

Our period search analysis also reveals the occurrence of other periodic signals in the amplitude of the small-scale variability. The peak at $\sim$6.6 d is consistent with the reported rotation period of the star by \citet{Biddle2018}. It is expected that accretion hot spots show some level of periodicity in line with the star's rotation because accreting material at the co-rotation radius orbits at the same rate. Material accreting from the co-rotation radius should fall on the same location on the stellar surface, even as it rotates. Recently, \citet{Donati2020} presented spectropolarimetric data that point to the $\sim$9.0 d signal as the star's rotation. They find a $\sim$9.0 d periodic variation in the Narrow Core He I D$_{3}$ line complex, which they also attribute to a periodic accretion signature. Rotational modulation is also detected via cool spots on the stellar surface when they come into and out of view as the star rotates. However in the case of young stars, cool spots are not known to change on rapid timescales like those identified in the small-amplitude variability; instead, they can remain relatively constant for decades at a time \citep{Stelzer2003,Carvalho2020}. The strength of the $\sim$6.6 d and $\sim$9.0 d periods identified in the amplitude of the small-scale variability lightcurve (Figure \ref{fig:crosscuts}) is comparable to that seen in the large-scale variability (Figure \ref{fig:lc}), which may signify that, in the case of CI Tau, the effect of hotspots on the photometric rotation signal may dominate rotational modulation compared to the effect of cool spots. Alternatively, this may indicate that hotspots occur where cool spots don't; the presence of cool spots decreases the total brightness of the stellar disk which increases the apparent contrast of the hotspot accretion shocks.

\subsection*{P\MakeLowercase{eriods beyond 9 d}}\label{sec:3.1}
There are also peaks in both periodograms that correspond to neither the star's rotation ($\sim$6.6 d) nor the planet's orbit ($\sim$9.0 d). These peaks also do not correspond with integer factors of the star's rotation period and planet's orbital period, suggesting that there are other mechanisms affecting the observed periodicity in accretion signatures. These peaks are located at $\sim$11.5 d, $\sim$14.2 d, and $\sim$20.0-24.4 d. The strongest peak is located in the range of $\sim$20.0-24.4 d. Despite its strong power, this period covers a large fraction of the duration of observations, barely cycling 3 full periods. We tested the possibility of this being the synodic period of any combination of the other signals and it does agree with the synodic period of the $\sim$6.6 d and $\sim$9.0 d signals, and the $\sim$14.2 d and $\sim$9.0 d signals. However tests of combined sine functions with 6.6 and 9.0 d periods as well as 14.2 and 9.0 d periods with identical sampling to the \ktwo data did not produce a similar peak as the $\sim$20.0-24.4 d one seen in CI Tau. This is likely not the synodic period. The weakest peak corresponds to a period of $\sim$11.5 d. This period lies within half the 20.0-24.4 d period range, likely explaining it as an alias, rather than a true period detection. The $\sim$14.2 d period does not appear to be an alias in the periodogram and there are no known \ktwo systematics occur periodically on these timescales.

One possible physical explanation for these signals may originate from pulsed accretion by a planetary body, much like CI Tau b. Main sequence stars with a hot Jupiter are not observed to host additional planets with periods within a factor of a few times that of the hot Jupiter \citep{Steffen2012}; however, it is possible that young stars may exhibit different planetary architectures compared to mature systems (e.g., planets may be undergoing active migration).

\section{Summary} \label{sec:summary}
Our analysis of the small-amplitude variability in the \ktwo lightcurve of CI Tau shows that although the brightness fluctuations are stochastic, the strength of the amplitude of these fluctuations varies periodically in time. We isolate the small-scale photometric variability in the \ktwo photometry by applying a Gaussian Process model to capture the large-scale variability. Our period-search of the time-dependent variance of the scatter of the residuals reveals peaks at the same periods identified in the large-scale variability. This is likely the cause of a stochastic bursting effect which can happen as a result of density changes at the inner edge of the disk on timescales of about a day \citet{Robinson2017}. Such density perturbations may be induced by dense, inhomogeneous regions orbiting the star. CI Tau is known to host a hot Jupiter near the inner-edge of the disk \citep{Johns-Krull2016, Flagg2019} and \citet{Teyssandier&Lai2019a} show that a hot Jupiter similar to CI Tau b can affect the density of the disk material where it orbits (and therefore the accretion rate), causing pulsed accretion detectable in time-series photometry as was observed by \citet{Biddle2018}. The rotation period of the star also appears in the periodogram of the small-scale variability, possibly suggesting that in the case of CI Tau, hotspots at the foot of accretion streams that are magnetically locked with the disk may dominate the contribution of rotation modulation over cool spots. Periodic signals corresponding to Keplerian distances beyond the orbit of the known planet also appear in the time-dependent amplitude of the small-scale variability. The origin of these periodic signals is unclear, though it remains possible that they could also be a result of pulsed accretion by other planet-mass companions, indicative possibly of dense planet-packing in unstable, young system architectures.

\section*{Acknowledgements} \label{sec:acknowledgements}
We thank the anonymous referee for a prompt report and insightful comments and suggestions, which have improved this manuscript. This work analyzes data collected by the \ktwo mission. Funding for the Kepler mission is provided by the NASA Science Mission directorate. L.I.B., L.A.P., and J.L. acknowledge support from NASA through an Astrophysics Data Analysis Program grant to Lowell Observatory (grant 80NSSC20K1001). ACC and MMJ acknowledge support from
the Science and Technology Facilities Council (STFC) consolidated
grant number ST/R000824/1, and the support of the visiting scientist program at Lowell Observatory in January 2019 and January 2020. Data were obtained using the Mikulski Archive for Space Telescopes (MAST). STScI is operated by the Association of Universities for Research in Astronomy, Inc., under NASA contract NAS5-26555. Support for MAST for non-HST data is provided by the NASA Office of Space Science via grant NNX09AF08G and by other grants and contracts. This project made use of NASA's Astrophysics Data System Bibliographic Services and the SIMBAD database, operated at CDS, Strasbourg, France.


\bibliographystyle{aasjournal}
\bibliography{bibtexfile63}{}

\begin{thebibliography}{}
\expandafter\ifx\csname natexlab\endcsname\relax\def\natexlab#1{#1}\fi
\providecommand{\url}[1]{\href{#1}{#1}}
\providecommand{\dodoi}[1]{doi:~\href{http://doi.org/#1}{\nolinkurl{#1}}}
\providecommand{\doeprint}[1]{\href{http://ascl.net/#1}{\nolinkurl{http://ascl.net/#1}}}
\providecommand{\doarXiv}[1]{\href{https://arxiv.org/abs/#1}{\nolinkurl{https://arxiv.org/abs/#1}}}

\bibitem[{{Aigrain} {et~al.}(2015){Aigrain}, {Hodgkin}, {Irwin}, {Lewis}, \&
  {Roberts}}]{aigrain2015}
{Aigrain}, S., {Hodgkin}, S.~T., {Irwin}, M.~J., {Lewis}, J.~R., \& {Roberts},
  S.~J. 2015, \mnras, 447, 2880, \dodoi{10.1093/mnras/stu2638}

\bibitem[{{Aigrain} {et~al.}(2016){Aigrain}, {Parviainen}, \&
  {Pope}}]{aigrain2016}
{Aigrain}, S., {Parviainen}, H., \& {Pope}, B.~J.~S. 2016, \mnras, 459, 2408,
  \dodoi{10.1093/mnras/stw706}

\bibitem[{{Aigrain} {et~al.}(2012){Aigrain}, {Pont}, \& {Zucker}}]{Aigrain2012}
{Aigrain}, S., {Pont}, F., \& {Zucker}, S. 2012, \mnras, 419, 3147,
  \dodoi{10.1111/j.1365-2966.2011.19960.x}

\bibitem[{{Biddle} {et~al.}(2018){Biddle}, {Johns-Krull}, {Llama}, {Prato}, \&
  {Skiff}}]{Biddle2018}
{Biddle}, L.~I., {Johns-Krull}, C.~M., {Llama}, J., {Prato}, L., \& {Skiff},
  B.~A. 2018, \apjl, 853, L34, \dodoi{10.3847/2041-8213/aaa897}

\bibitem[{{Calvet} \& {Hartmann}(1992)}]{Calvet1992}
{Calvet}, N., \& {Hartmann}, L. 1992, \apj, 386, 239, \dodoi{10.1086/171010}

\bibitem[{{Carvalho et al.}(Submitted 2020)}]{Carvalho2020}
{Carvalho et al.} Submitted 2020, \apj

\bibitem[{{Chauvin} {et~al.}(2004){Chauvin}, {Lagrange}, {Dumas}, {Zuckerman},
  {Mouillet}, {Song}, {Beuzit}, \& {Lowrance}}]{Chauvin2004}
{Chauvin}, G., {Lagrange}, A.~M., {Dumas}, C., {et~al.} 2004, \aap, 425, L29,
  \dodoi{10.1051/0004-6361:200400056}

\bibitem[{{Chauvin} {et~al.}(2005){Chauvin}, {Lagrange}, {Dumas}, {Zuckerman},
  {Mouillet}, {Song}, {Beuzit}, \& {Lowrance}}]{Chauvin2005}
---. 2005, \aap, 438, L25, \dodoi{10.1051/0004-6361:200500116}

\bibitem[{{Clarke} {et~al.}(2018){Clarke}, {Tazzari}, {Juhasz}, {Rosotti},
  {Booth}, {Facchini}, {Ilee}, {Johns-Krull}, {Kama}, {Meru}, \&
  {Prato}}]{Clarke2018}
{Clarke}, C.~J., {Tazzari}, M., {Juhasz}, A., {et~al.} 2018, \apjl, 866, L6,
  \dodoi{10.3847/2041-8213/aae36b}

\bibitem[{{Cohen} {et~al.}(2004){Cohen}, {Herbst}, \& {Williams}}]{Cohen2004}
{Cohen}, R.~E., {Herbst}, W., \& {Williams}, E.~C. 2004, \aj, 127, 1602,
  \dodoi{10.1086/381925}

\bibitem[{{Costigan} {et~al.}(2014){Costigan}, {Vink}, {Scholz}, {Ray}, \&
  {Testi}}]{Costigan2014}
{Costigan}, G., {Vink}, J.~S., {Scholz}, A., {Ray}, T., \& {Testi}, L. 2014,
  \mnras, 440, 3444, \dodoi{10.1093/mnras/stu529}

\bibitem[{{Currie} {et~al.}(2019){Currie}, {Marois}, {Cieza}, {Mulders},
  {Lawson}, {Caceres}, {Rodriguez-Ruiz}, {Wisniewski}, {Guyon}, {Brandt},
  {Kasdin}, {Groff}, {Lozi}, {Chilcote}, {Hodapp}, {Jovanovic}, {Martinache},
  {Skaf}, {Lyra}, {Tamura}, {Asensio-Torres}, {Dong}, {Grady}, {Gerard},
  {Fukagawa}, {Hand}, {Hayashi}, {Henning}, {Kudo}, {Kuzuhara}, {Kwon},
  {McElwain}, \& {Uyama}}]{Currie2019}
{Currie}, T., {Marois}, C., {Cieza}, L., {et~al.} 2019, \apjl, 877, L3,
  \dodoi{10.3847/2041-8213/ab1b42}

\bibitem[{{Damasso} {et~al.}(2020){Damasso}, {Lanza}, {Benatti}, {Rajpaul},
  {Mallonn}, {Desidera}, {Biazzo}, {D'Orazi}, {Malavolta}, {Nardiello},
  {Rainer}, {Borsa}, {Affer}, {Bignamini}, {Bonomo}, {Carleo}, {Claudi},
  {Cosentino}, {Covino}, {Giacobbe}, {Gratton}, {Harutyunyan}, {Knapic},
  {Leto}, {Maggio}, {Maldonado}, {Mancini}, {Micela}, {Molinari}, {Nascimbeni},
  {Pagano}, {Piotto}, {Poretti}, {Scandariato}, {Sozzetti}, {Capuzzo Dolcetta},
  {Di Mauro}, {Carosati}, {Fiorenzano}, {Frustagli}, {Pedani}, {Pinamonti},
  {Stoev}, \& {Turrini}}]{Damasso2020}
{Damasso}, M., {Lanza}, A.~F., {Benatti}, S., {et~al.} 2020, arXiv e-prints,
  arXiv:2008.09445.
\newblock \doarXiv{2008.09445}

\bibitem[{{David} {et~al.}(2019{\natexlab{a}}){David}, {Petigura}, {Luger},
  {Foreman-Mackey}, {Livingston}, {Mamajek}, \& {Hillenbrand}}]{David2019a}
{David}, T.~J., {Petigura}, E.~A., {Luger}, R., {et~al.} 2019{\natexlab{a}},
  \apjl, 885, L12, \dodoi{10.3847/2041-8213/ab4c99}

\bibitem[{{David} {et~al.}(2016){David}, {Hillenbrand}, {Petigura},
  {Carpenter}, {Crossfield}, {Hinkley}, {Ciardi}, {Howard}, {Isaacson}, {Cody},
  {Schlieder}, {Beichman}, \& {Barenfeld}}]{david2016}
{David}, T.~J., {Hillenbrand}, L.~A., {Petigura}, E.~A., {et~al.} 2016, \nat,
  534, 658, \dodoi{10.1038/nature18293}

\bibitem[{{David} {et~al.}(2019{\natexlab{b}}){David}, {Cody}, {Hedges},
  {Mamajek}, {Hillenbrand}, {Ciardi}, {Beichman}, {Petigura}, {Fulton},
  {Isaacson}, {Howard}, {Gagn{\'e}}, {Saunders}, {Rebull}, {Stauffer},
  {Vasisht}, \& {Hinkley}}]{David2019b}
{David}, T.~J., {Cody}, A.~M., {Hedges}, C.~L., {et~al.} 2019{\natexlab{b}},
  \aj, 158, 79, \dodoi{10.3847/1538-3881/ab290f}

\bibitem[{{Donati} {et~al.}(2017){Donati}, {Yu}, {Moutou}, {Cameron}, {Malo},
  {Grankin}, {H{\'e}brard}, {Hussain}, {Vidotto}, {Alencar}, {Haywood},
  {Bouvier}, {Petit}, {Takami}, {Herczeg}, {Gregory}, {Jardine}, {Morin}, \&
  {MaTYSSE Collaboration}}]{donati2017}
{Donati}, J.~F., {Yu}, L., {Moutou}, C., {et~al.} 2017, \mnras, 465, 3343,
  \dodoi{10.1093/mnras/stw2904}

\bibitem[{{Donati} {et~al.}(2020){Donati}, {Bouvier}, {Alencar}, {Moutou},
  {Malo}, {Takami}, {M{\'e}nard}, {Dougados}, {Hussain}, \& {The Matysse
  Collaboration}}]{Donati2020}
{Donati}, J.~F., {Bouvier}, J., {Alencar}, S.~H., {et~al.} 2020, \mnras, 491,
  5660, \dodoi{10.1093/mnras/stz3368}

\bibitem[{{Feinstein} {et~al.}(2020){Feinstein}, {Montet}, {Ansdell}, {Nord},
  {Bean}, {G{\"u}nther}, {Gully-Santiago}, \& {Schlieder}}]{Feinstein2020}
{Feinstein}, A.~D., {Montet}, B.~T., {Ansdell}, M., {et~al.} 2020, arXiv
  e-prints, arXiv:2005.07710.
\newblock \doarXiv{2005.07710}

\bibitem[{{Flagg} {et~al.}(2019){Flagg}, {Johns-Krull}, {Nofi}, {Llama},
  {Prato}, {Sullivan}, {Jaffe}, \& {Mace}}]{Flagg2019}
{Flagg}, L., {Johns-Krull}, C.~M., {Nofi}, L., {et~al.} 2019, \apjl, 878, L37,
  \dodoi{10.3847/2041-8213/ab276d}

\bibitem[{{Foreman-Mackey} {et~al.}(2017){Foreman-Mackey}, {Agol},
  {Ambikasaran}, \& {Angus}}]{Foreman-Mackey2017}
{Foreman-Mackey}, D., {Agol}, E., {Ambikasaran}, S., \& {Angus}, R. 2017, \aj,
  154, 220, \dodoi{10.3847/1538-3881/aa9332}

\bibitem[{{Foreman-Mackey} {et~al.}(2019){Foreman-Mackey}, {Barentsen}, \&
  {Barclay}}]{Foreman-Mackey2019}
{Foreman-Mackey}, D., {Barentsen}, G., \& {Barclay}, T. 2019, {dfm/exoplanet:
  exoplanet v0.1.6}, v0.1.6,  Zenodo, \dodoi{10.5281/zenodo.2651251}

\bibitem[{{Fromang} \& {Papaloizou}(2006)}]{Fromang2006}
{Fromang}, S., \& {Papaloizou}, J. 2006, \aap, 452, 751,
  \dodoi{10.1051/0004-6361:20054612}

\bibitem[{{Haywood} {et~al.}(2014){Haywood}, {Collier Cameron}, {Queloz},
  {Barros}, {Deleuil}, {Fares}, {Gillon}, {Lanza}, {Lovis}, {Moutou}, {Pepe},
  {Pollacco}, {Santerne}, {S{\'e}gransan}, \& {Unruh}}]{Haywood2014}
{Haywood}, R.~D., {Collier Cameron}, A., {Queloz}, D., {et~al.} 2014, \mnras,
  443, 2517, \dodoi{10.1093/mnras/stu1320}

\bibitem[{{Herbst} {et~al.}(2007){Herbst}, {Eisl{\"o}ffel}, {Mundt}, \&
  {Scholz}}]{Herbst2007}
{Herbst}, W., {Eisl{\"o}ffel}, J., {Mundt}, R., \& {Scholz}, A. 2007, in
  Protostars and Planets V, ed. B.~{Reipurth}, D.~{Jewitt}, \& K.~{Keil}, 297.
\newblock \doarXiv{astro-ph/0603673}

\bibitem[{{Herbst} {et~al.}(1994){Herbst}, {Herbst}, {Grossman}, \&
  {Weinstein}}]{Herbst1994}
{Herbst}, W., {Herbst}, D.~K., {Grossman}, E.~J., \& {Weinstein}, D. 1994, \aj,
  108, 1906, \dodoi{10.1086/117204}

\bibitem[{{Huerta} {et~al.}(2008){Huerta}, {Johns-Krull}, {Prato}, {Hartigan},
  \& {Jaffe}}]{Huerta2008}
{Huerta}, M., {Johns-Krull}, C.~M., {Prato}, L., {Hartigan}, P., \& {Jaffe},
  D.~T. 2008, \apj, 678, 472, \dodoi{10.1086/526415}

\bibitem[{{Ivezi{\'c}} {et~al.}(2014){Ivezi{\'c}}, {Connelly}, {Vand erPlas},
  \& {Gray}}]{Ivezic2014}
{Ivezi{\'c}}, {\v{Z}}., {Connelly}, A.~J., {Vand erPlas}, J.~T., \& {Gray}, A.
  2014, {Statistics, Data Mining, and Machine Learning in Astronomy}

\bibitem[{{Jenkins} {et~al.}(2010){Jenkins}, {Caldwell}, {Chandrasekaran},
  {Twicken}, {Bryson}, {Quintana}, {Clarke}, {Li}, {Allen}, {Tenenbaum}, {Wu},
  {Klaus}, {Middour}, {Cote}, {McCauliff}, {Girouard}, {Gunter}, {Wohler},
  {Sommers}, {Hall}, {Uddin}, {Wu}, {Bhavsar}, {Van Cleve}, {Pletcher},
  {Dotson}, {Haas}, {Gilliland}, {Koch}, \& {Borucki}}]{jenkins2010}
{Jenkins}, J.~M., {Caldwell}, D.~A., {Chandrasekaran}, H., {et~al.} 2010,
  \apjl, 713, L87, \dodoi{10.1088/2041-8205/713/2/L87}

\bibitem[{{Johns-Krull} {et~al.}(2016){Johns-Krull}, {McLane}, {Prato},
  {Crockett}, {Jaffe}, {Hartigan}, {Beichman}, {Mahmud}, {Chen}, {Skiff},
  {Cauley}, {Jones}, \& {Mace}}]{Johns-Krull2016}
{Johns-Krull}, C.~M., {McLane}, J.~N., {Prato}, L., {et~al.} 2016, \apj, 826,
  206, \dodoi{10.3847/0004-637X/826/2/206}

\bibitem[{{Koenigl}(1991)}]{Koenigl1991}
{Koenigl}, A. 1991, \apjl, 370, L39, \dodoi{10.1086/185972}

\bibitem[{{Kraus} \& {Ireland}(2012)}]{kraus2012}
{Kraus}, A.~L., \& {Ireland}, M.~J. 2012, \apj, 745, 5,
  \dodoi{10.1088/0004-637X/745/1/5}

\bibitem[{{Mann} {et~al.}(2016){Mann}, {Newton}, {Rizzuto}, {Irwin}, {Feiden},
  {Gaidos}, {Mace}, {Kraus}, {James}, {Ansdell}, {Charbonneau}, {Covey},
  {Ireland}, {Jaffe}, {Johnson}, {Kidder}, \& {Vanderburg}}]{mann2016}
{Mann}, A.~W., {Newton}, E.~R., {Rizzuto}, A.~C., {et~al.} 2016, \aj, 152, 61,
  \dodoi{10.3847/0004-6256/152/3/61}

\bibitem[{{Orlando} {et~al.}(2010){Orlando}, {Sacco}, {Argiroffi}, {Reale},
  {Peres}, \& {Maggio}}]{Orlando2010}
{Orlando}, S., {Sacco}, G.~G., {Argiroffi}, C., {et~al.} 2010, \aap, 510, A71,
  \dodoi{10.1051/0004-6361/200913565}

\bibitem[{{Press} {et~al.}(1992){Press}, {Teukolsky}, {Vetterling}, \&
  {Flannery}}]{Press1992}
{Press}, W.~H., {Teukolsky}, S.~A., {Vetterling}, W.~T., \& {Flannery}, B.~P.
  1992, {Numerical recipes in C. The art of scientific computing}

\bibitem[{{Queloz} {et~al.}(2001){Queloz}, {Henry}, {Sivan}, {Baliunas},
  {Beuzit}, {Donahue}, {Mayor}, {Naef}, {Perrier}, \& {Udry}}]{Queloz2001}
{Queloz}, D., {Henry}, G.~W., {Sivan}, J.~P., {et~al.} 2001, \aap, 379, 279,
  \dodoi{10.1051/0004-6361:20011308}

\bibitem[{{Robinson} {et~al.}(2017){Robinson}, {Owen}, {Espaillat}, \&
  {Adams}}]{Robinson2017}
{Robinson}, C.~E., {Owen}, J.~E., {Espaillat}, C.~C., \& {Adams}, F.~C. 2017,
  \apj, 838, 100, \dodoi{10.3847/1538-4357/aa671f}

\bibitem[{{Romanova} {et~al.}(2012){Romanova}, {Ustyugova}, {Koldoba}, \&
  {Lovelace}}]{Romanova2012}
{Romanova}, M.~M., {Ustyugova}, G.~V., {Koldoba}, A.~V., \& {Lovelace},
  R.~V.~E. 2012, \mnras, 421, 63, \dodoi{10.1111/j.1365-2966.2011.20055.x}

\bibitem[{{Shu} {et~al.}(1994){Shu}, {Najita}, {Ruden}, \& {Lizano}}]{Shu1994}
{Shu}, F.~H., {Najita}, J., {Ruden}, S.~P., \& {Lizano}, S. 1994, \apj, 429,
  797, \dodoi{10.1086/174364}

\bibitem[{{Steffen} {et~al.}(2012){Steffen}, {Ragozzine}, {Fabrycky}, {Carter},
  {Ford}, {Holman}, {Rowe}, {Welsh}, {Borucki}, {Boss}, {Ciardi}, \&
  {Quinn}}]{Steffen2012}
{Steffen}, J.~H., {Ragozzine}, D., {Fabrycky}, D.~C., {et~al.} 2012,
  Proceedings of the National Academy of Science, 109, 7982,
  \dodoi{10.1073/pnas.1120970109}

\bibitem[{{Stelzer} {et~al.}(2003){Stelzer}, {Fern{\'a}ndez}, {Costa},
  {Gameiro}, {Grankin}, {Henden}, {Guenther}, {Mohanty}, {Flaccomio},
  {Burwitz}, {Jayawardhana}, {Predehl}, \& {Durisen}}]{Stelzer2003}
{Stelzer}, B., {Fern{\'a}ndez}, M., {Costa}, V.~M., {et~al.} 2003, \aap, 411,
  517, \dodoi{10.1051/0004-6361:20031414}

\bibitem[{{Teyssandier} \& {Lai}(2019)}]{Teyssandier&Lai2019a}
{Teyssandier}, J., \& {Lai}, D. 2019, arXiv e-prints, arXiv:1911.08492.
\newblock \doarXiv{1911.08492}

\bibitem[{{Teyssandier} {et~al.}(2019){Teyssandier}, {Lai}, \&
  {Vick}}]{Teyssandier2019}
{Teyssandier}, J., {Lai}, D., \& {Vick}, M. 2019, \mnras, 486, 2265,
  \dodoi{10.1093/mnras/stz1011}

\bibitem[{{Twicken} {et~al.}(2010){Twicken}, {Chandrasekaran}, {Jenkins},
  {Gunter}, {Girouard}, \& {Klaus}}]{Twicken2010}
{Twicken}, J.~D., {Chandrasekaran}, H., {Jenkins}, J.~M., {et~al.} 2010, in
  Society of Photo-Optical Instrumentation Engineers (SPIE) Conference Series,
  Vol. 7740, Software and Cyberinfrastructure for Astronomy, ed. N.~M.
  {Radziwill} \& A.~{Bridger}, 77401U, \dodoi{10.1117/12.856798}

\bibitem[{{Yu} {et~al.}(2017){Yu}, {Donati}, {H{\'e}brard}, {Moutou}, {Malo},
  {Grankin}, {Hussain}, {Collier Cameron}, {Vidotto}, {Baruteau}, {Alencar},
  {Bouvier}, {Petit}, {Takami}, {Herczeg}, {Gregory}, {Jardine}, {Morin},
  {M{\'e}nard}, \& {Matysse Collaboration}}]{yu2017}
{Yu}, L., {Donati}, J.~F., {H{\'e}brard}, E.~M., {et~al.} 2017, \mnras, 467,
  1342, \dodoi{10.1093/mnras/stx009}

\bibitem[{{Zechmeister} \& {K{\"u}rster}(2009)}]{Zechmeister2009}
{Zechmeister}, M., \& {K{\"u}rster}, M. 2009, \aap, 496, 577,
  \dodoi{10.1051/0004-6361:200811296}

\end{thebibliography}

\appendix
\section{Appendix} \label{sec:appendix}

\citet{Foreman-Mackey2017} presented a framework for directly and accurately computing a specific type of GP that scales linearly with the number of data points, i.e., as $\mathcal{O}(N)$. Known as \textit{celerite}, the framework requires the covariance function to be represented by a mixture of exponentials but does not require the data to be evenly spaced or to be associated with uniform uncertainties. 

\begin{figure*}[h!]
\centering
    \includegraphics[width=\textwidth]{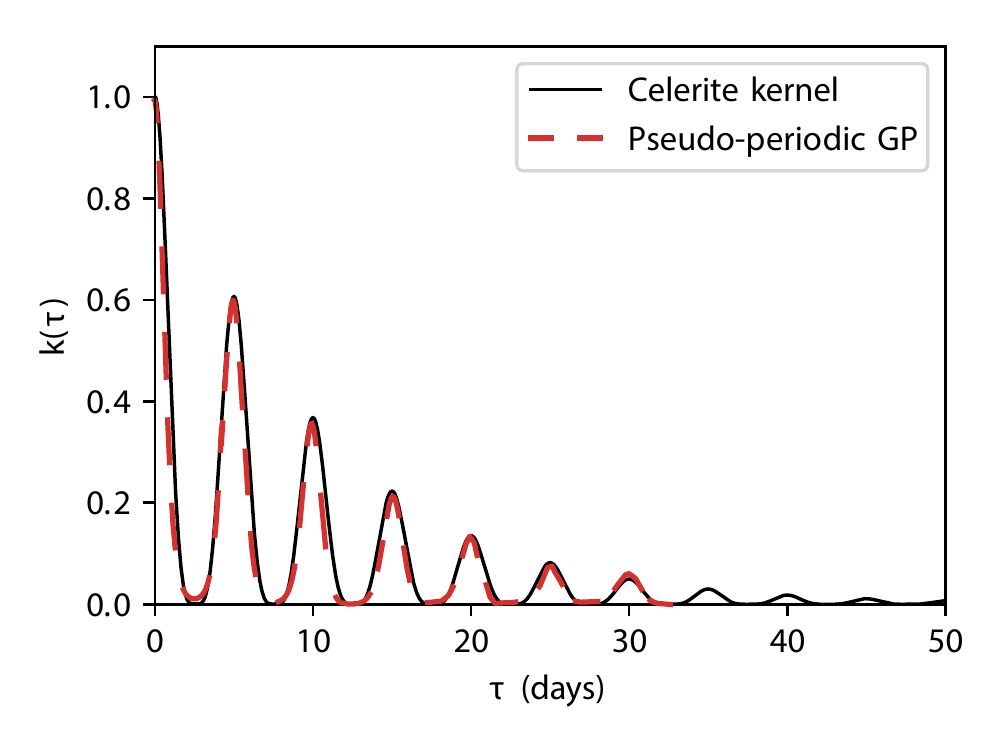}
    \caption{Our {\it celerite} kernel captures the structure of a pseudo-periodic Gaussian Process kernel and is computationally quick and efficient, enabling us to execute it on many \ktwo light curves.}
\label{fig:kernel}
\end{figure*}

The presence of starspots and other magnetic features such as faculae and plage on the stellar surface manifest as quasi-periodic variability in the light curve. Typically, this variability is modeled using a pseudo-periodic covariance function of the form 
\begin{equation}
    k(\tau) = \theta_1^2\exp\left[-\frac{\tau^2}{\theta_2^2} - \frac{\sin^2\left(\frac{\pi\tau}{\theta_3}\right)}{\theta_4^2}\right],
\end{equation}
where, $\theta_1$ is the amplitude of the GP, $\theta_2$ is the recurrence time-scale, $\theta_3$ is the decay time-scale, which is directly related to the star spot lifetime, and $\theta_4$ is a smoothing-parameter \citep{Aigrain2012,Haywood2014}. To model such variability with \textit{celerite}, we have constructed a covariance function with the following form, 
\begin{eqnarray}
k(\tau) = \textrm{e}^{-c\tau}\left[a\cos(d\,\tau) + b\sin(d\,\tau) + e\cos(2d\,\tau) +f\sin(2d\,\tau) +g \right],
\label{eqn:celerite_kernel_sines}
\end{eqnarray}
or equivalently, 
\begin{eqnarray}
\nonumber    k(\tau) &=& \frac{1}{2}\left[(a+ib)\textrm{e}^{-(c+id)\tau} + (a-ib)\textrm{e}^{-(c-id)\tau}\right]  \\ 
           \nonumber&+& \frac{1}{2}\left[(e+if)\textrm{e}^{-(c+2id)\tau} + (e-if)\textrm{e}^{-(c-2id)\tau}\right] \\
           &+& g\textrm{e}^{-c\tau}.
\label{eqn:celerite_kernel_exp}
\end{eqnarray}
To solve for the coefficients $(a,b,c,d,e,f,g)$, we impose the following constraints on $k(\tau)$: $k(0)=1, k(\pi/d)=0, k'(0)=0,k'(\pi/d)=0,k'(2\pi/d)=0$. These constraints allow us to solve for the coefficients in terms of the physically meaningful quantities: light curve amplitude $(A)$, spot lifetime $(\tau_{\rm spot})$, and rotation period $(P_{\rm rot})$: 
\begin{align*}
a &= \frac{A}{2}, & b &= \frac{A}{2}\frac{c}{d}, & c &=1 / \tau_{\rm spot}, &d &= 2\pi /P_{\rm rot}, & 
e &= \frac{A}{8}, & f &= \frac{A}{4}\frac{c}{d},& &g &= \epsilon\,A,
\end{align*}
where, $\epsilon$ is a small perturbation. Figure \ref{fig:kernel} shows a comparison between the traditional pseudo-periodic GP kernel (Equation \ref{eqn:celerite_kernel_sines}) and our \textit{celerite} kernel (Equation \ref{eqn:celerite_kernel_exp}) for a star with $P_{\rm rot} =5$ days, $\tau_{\rm spot} = 10\times\,P_{\rm rot}$, and $A=1$.  The covariance function maintains the properties of the pseudo-periodic GP, peaking near the stellar rotation period (similar longitudes on the stellar surface), and is 0 at multiples of half of the stellar rotation period (i.e., the opposite side of the star).

\end{document}